\documentclass{amos}

\usepackage{amsmath}
\usepackage{amssymb}
\usepackage{xcolor}
\usepackage[numbers]{natbib}
\usepackage[hidelinks]{hyperref}

\title{\TitleFont Star-Galaxy Separation via Gaussian Processes with Model Reduction}

\author{
	Im\`ene R. Goumiri \\ Physics Division, Lawrence Livermore National Laboratory 
\and\vspace{0.5em}
	Amanda L. Muyskens \\ Engineering Division, Lawrence Livermore National Laboratory 
\and
	Michael D. Schneider \\ Physics Division, Lawrence Livermore National Laboratory 
\and
	Benjamin W. Priest \\ Center for Applied Scientific Computing, Lawrence Livermore National Laboratory 
\and
	Robert E. Armstrong \\ Physics Division, Lawrence Livermore National Laboratory 
}

\date{}

\begin{document} 

\maketitle

\vspace{-0.8in}
\begin{center}
LLNL-CONF-813954
\end{center}
\vspace{0.5in}

\begin{abstract}\normalsize

Modern cosmological surveys such as the Hyper Suprime-Cam (HSC) survey produce a huge volume of low-resolution images of both distant galaxies and dim stars in our own galaxy.  Being able to automatically classify these images is a long-standing problem in astronomy and critical to a number of different scientific analyses.
Recently, the challenge of ``star-galaxy'' classification has been approached with Deep Neural Networks (DNNs), which are good at learning complex nonlinear embeddings. 
However, DNNs are known to overconfidently extrapolate on unseen data and require a large volume of training images that accurately capture the data distribution to be considered reliable. 
Gaussian Processes (GPs), which infer posterior distributions over functions and naturally quantify uncertainty, haven’t been a tool of choice for this task mainly because popular kernels exhibit limited expressivity on complex and high-dimensional data.

In this paper, we present a novel approach to the star-galaxy separation problem that uses GPs and reap their benefits while solving many of the issues traditionally affecting them for classification of high-dimensional celestial image data. 
After an initial filtering of the raw data of star and galaxy image cutouts, we first reduce the dimensionality of the input images by using a Principal Components Analysis (PCA) before applying GPs using a simple Radial Basis Function (RBF) kernel on the reduced data. 
Using this method, we greatly improve the accuracy of the classification over a basic application of GPs while improving the computational efficiency and scalability of the method.
	
\end{abstract}


\section{Introduction} \label{sec:intro}

Modern telescopes have enabled the capture of heretofore undetectable faint stars and galaxies.  Recent large-scale surveys, like the Dark Energy Survey (DES)~\cite{des2005,abbott2018} and the Hyper Surpime-Cam (HSC)~\cite{miyazaki2018,Aihara17} survey, have generated vast quantities of data at increasing depths, requiring the development of automated algorithms to distinguish stars from galaxies. 
For bright objects the problem is usually fairly straightforward; however for fainter objects the problem becomes more difficult.  
At the depths of modern ground-based surveys faint galaxies are almost indistinguishable from point-like sources. 
Observations from space, where the resolution is better can significantly improve classification, but telescope time is too costly to observe large regions of the sky. 
Another factor is that the number of stars quickly decreases as a function of magnitude.

Previous studies have applied several different approaches to star-galaxy separation.  The morphological approach~\cite{Vasconcellos11, Slater20} uses the difference in shape in comparison with the point-spread function (PSF). This approach typically uses summary information to reduce the complexity of the problem as the number of pixels for different filters and PSF can be high-dimensional.  A similar morphological approach is used in both DES and HSC where a comparison of the flux ratio of a galaxy model to a model using the PSF is used. An alternative approach uses the fact that stars and galaxies have different spectral energy distributions.  This translates into stars and galaxies clustering into different regions of color space.  For example, \citet{Pollo10} used selections in color-color space for separation in far-infrared data.

There is also a rich literature of feature based machine learning solutions using color and morphological information as input into decision trees~\cite{vasconcellos2011decision, sevilla2015effect}, neural networks~\cite{odewahn1992} and ensemble methods such as random forests~\cite{kim2015hybrid}.  Although decision tree-based models offer convenient interpretability features, they often generalize poorly and are sensitive to crafted feature vectors, usually based upon expert knowledge in applications.  Given the high dimensionality of the data to be processed and the difficulty encountered in capturing generalizable data features using expert knowledge, it is tempting to instead \emph{learn} an appropriate feature representation.
Representation learning models include deep neural networks (DNNs) and kernel methods such as support vector machines (SVMs) and Gaussian processes (GPs).
Indeed, \citet{odewahn1992automated} applied shallow neural networks to the star--galaxy separation problem as early as 1992.
Recently, deep convolutional neural networks (CNNs) have achieved success in many areas of image classification due to their success in capturing local interactions in structured pixel data~\cite{krizhevsky2012imagenet}.
\citet{kim2016star} recently utilized CNNs to demonstrate star-galaxy separation performance that was competitive with a conventional random forest-based classifier~\cite{kim2015hybrid}.

In spite of their surprising generalization success in many problems, DNNs tend to poorly quantify their own uncertainty~\cite{blundell2015weight, gal2016dropout}.
Importantly, this leads to overconfidently extrapolating on data that is far from training examples~\cite{guo2017calibration}.
Overconfident extrapolation presents a significant problem to an automated classification pipeline, where one desires to detect low-confidence predictions for possible human intervention.
For example, the removal of ambiguous sources to reduce systematic biases is a major issue facing the statistical analysis of astrophysical populations.
Practitioners often attempt to ameliorate the generalization gap by way of data augmentation and the use of very large labeled datasets that covers the full data distribution.
However, both mitigations introduce additional cost and implementation complexity when training a model. 
For many astronomical applications, it is also not feasible to simulate unbiased and complete training data distributions because the distributions are unknown (indeed, learning these distributions is a primary scientific objective of many surveys).

Kernel-based representations learning methods present a nonparametric alternative to DNNs.
A kernel learning model poses a possibly nonlinear feature map from the data space to a latent reproducing kernel Hilbert space and reasons over inner products in the latent space to make predictions.
Unlike DNNs, which explicitly construct their feature maps, kernel methods instead directly rely on the unique kernel function of their Hilbert space and do not compute their feature maps explicitly.
\citet{fadely2012star} demonstrated the application of simple SVMs to the star--galaxy separation problem. 
Gaussian processes (GPs) are a particularly attractive kernel model in many applications because they support fully-Bayesian inference, which naturally affords disciplined uncertainty quantification by way of examining the variance of the posterior distribution of a prediction. 
However, GPs have not yet been adopted in the star--galaxy separation literature, most likely because popular kernel functions are known to implicitly utilize feature maps that are less expressive than those produced by CNNs, particularly on high-dimensional image data~\cite{bradshaw2017adversarial}.

We demonstrate an application of GPs to the star--galaxy separation problem that ameliorates expressivity issues by first embedding high-dimensional image data onto a small number of their principle components using PCA.
We show that our augmented GP model demonstrates significant improvements over a na\"ive GP on cosmological survey data, even when using the simple RBF kernel.
We also demonstrate superlinear computational savings using PCA over the na\"ive application.
Our results motivate the serious consideration of GPs in star--galaxy separation as well as other large-scale cosmological machine learning problems.

\section{The data}

We consider data from the first public release of the HSC Subaru Strategic Program~\cite{Aihara17}.  
HSC is the deepest ongoing large-scale survey and provides an invaluable early testbed for algorithms being developed for the future Vera Rubin Observatory\footnote{\url{https://www.lsst.org/}}.
The HSC pipeline includes high level processing steps that generate coadded images and science-ready catalogs. More details on the software that was used to process the data can be found in~\cite{Bosch18}.

We create a training set of galaxy and star images from the UltraDeep COSMOS field.  
The COSMOS field is smaller in area but was observed many more times making the it significantly deeper than the main survey.  
It also overlaps space-based imaging from the Hubble Space Telescope (HST) where the resolution is much better.  
\citet{leauthaud2007cosmos} use HST images of the COSMOS field to separate stars and galaxies.  
The authors identify stars by looking for the stellar locus in the 2-d space of magnitude and peak surface brightness.  
They further claim their classification to be reliable up to magnitude $\sim25$.  
The classification breaks down at fainter magnitudes as point sources and galaxies appear identical.  
We cross match the HSC objects to the HST objects to label which object is a star or galaxy.  
The HSC data is quite a bit deeper ($\sim 27$  in the $i$ filter), so we must be careful in interpreting results at the faint end. 

\begin{figure}[!h]
  \centering
  \includegraphics[width=128px]{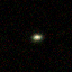}
  \includegraphics[width=128px]{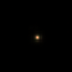}
  \caption{Example of images of a galaxy (left) and a star (right) from the HSC survey. Both images are composites where the RGB channels correspond to the $z$, $i$, and $r$ filters respectively. Other examples can be much harder to differentiate.}
  \label{fig:examples}
\end{figure}

We show in \autoref{fig:examples} two examples of images (galaxy vs star) obtained from the data.

\begin{figure}[!h]
  \centering
  \includegraphics{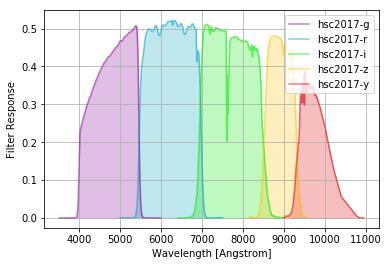}
  \caption{Different filters correspond to different frequency bands. For this analysis we used only images that had a signal to noise ratio greater than 10 in all bands.}
  \label{fig:filters}
\end{figure}

\def\fwhm{\ensuremath{\text{FWHM}_\text{PSF}}}
The HSC has five different filters, $g$, $r$, $i$, $z$, and $y$, corresponding to different frequency bands, as shown in \autoref{fig:filters}. The $y$ band data is not used because it is significantly shallower than the other bands.  We extract images for each filter from the deblended HSC images.  In an attempt to remove spurious sources, and to more closely match the HST catalogs, we only select objects with signal-to-noise ratio greater than 10 in all bands.  While this cut does remove some artifacts, we find that our catalogs still contains too many junk objects.  In addition to the signal-to-noise cut, we filter out images which fail the following test: the ratio of the $\chi^2$ value for a circular area of interest ($r < 2\,\fwhm$) over the chi squared value of an annulus region around it ($5\,\fwhm < r < 10\,\fwhm$) must be lower than 4, where \fwhm\ is the Full Width at Half Maximum (FWHM) of the Point Spread Function (PSF).

\section{Methodology}

Our goal is to build an automated star-galaxy separator which can be fed a set of images and can return the probability of being a star or galaxy. 
That separator first needs to be trained by learning from a set of pre-labelled data: the \emph{training set}. Then it must be validated against a different set of pre-labelled data: the \emph{test set}.

Our method consists of two parts: first we drastically reduce the dimension of the training set by using a Principal Components Analysis (PCA), then we train a Gaussian Process on the reduced-order data and apply it to produce predictions for test images.

\subsection{Model reduction} \label{sec:reduction}

For each observation, we produce a vector by concatenating the flattened images of the \emph{g}, \emph{r}, \emph{i}, and \emph{z} bands as well as their corresponding flattened PSF images. 
For a single band the image has $64 \times 64$ pixels and the PSF image has $43 \times 43$ pixels so the resulting vector of all bands has $23,780$ dimensions. 
Assuming $n$ training samples, the training set is a matrix of size $n \times 23780$. Similarly, if we have $m$ testing samples, we form a matrix of size $m \times 23780$. 
Define $W$ to be the $(n+m) \times  23780$ matrix whose first $n$ rows are the training data and whose last $m$ rows are the testing data.
Then, $W^T W$ is proportional to the sample covariance matrix of $W$, and we perform model reduction by the eigendecomposition of this matrix.

Define the eigendecomposition 
\begin{equation}
W^T W= P \Lambda P^T,
\end{equation}
where $\Lambda$ is a diagonal matrix that contains the sorted eigenvalues, and $P$ is a matrix that contains the corresponding eigenvectors. 
Define $P_{L}$ to be a $23780 \times L$ matrix comprised of columns of the first $L$ eigenvectors. 
Then the reduced dimension data is computed as $W P_L$.

\begin{figure}[!h]
	\includegraphics[width=\linewidth]{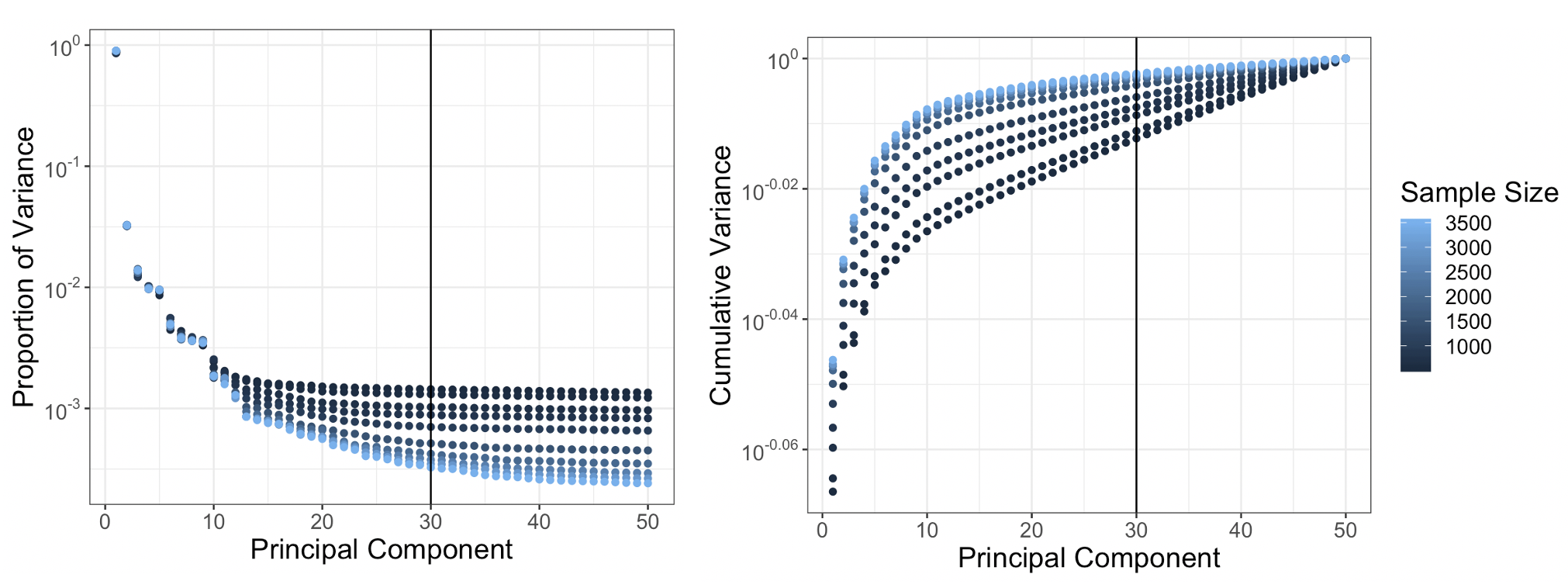}
	\caption{Proportion (left) and cumulative distribution (right) of variance in first 50 approximate principal components in a simulation iteration. For all tested sample sizes the cumulative variance explained in the first 30 components is more that 98\% of that contained in the first 50 components.}
	\label{fig:var}
\end{figure}

Practically speaking, since $W^T W$ has the dimension $23780 \times 23780$, its formation and eigendecomposition is intractable on a typical computer. 
Instead we approximately compute the largest eigenvalues and their corresponding eigenvectors using the methods in~\cite{baglama2005augmented}. 
This is implemented through the \texttt{partial\_eigen} function in the \texttt{irlba} package in R.
We found that 30 eigenvalues were sufficient to capture the most important features of the data and obtained a very good prediction accuracy as shown in \autoref{fig:var}. 
Further studies omitted here found similar results as our 30 components model for 50 and 100 eigenvectors. 
At the end of this operation, the reduced-order training set is now a matrix $X$ of size $n \times 30$, and the reduced-order testing set is now a matrix $X_*$ of size $m \times 30$.

\subsection{2-Class Gaussian Process Classification} \label{sec:classification}

GPs are flexible, nonparametric Bayesian models that specify a \emph{prior distribution over a function} $f : \mathcal{X} \rightarrow \mathcal{Y}$ that can be updated by data $\mathcal{D} \subset \mathcal{X} \times \mathcal{Y}$~\cite{williams2006gaussian}.
Coarsely, a GP is a collection of random variables, any finite subset of which has a multivariate Gaussian distribution.
We say that $f \sim \mathcal{GP}(0, k(\cdot, \cdot))$, where $k : \mathcal{X} \times \mathcal{X} \rightarrow \mathbb{R}$ is a kernel: a positive definite covariance function with hyperparameters $\theta$.

We will model the function $f$ mapping images embedded as described in \autoref{sec:reduction} to the scalar reals.
Thus, $\mathcal{X} = \mathbb{R}^{30}$ is the input domain, and $\mathcal{Y} = \mathbb{R}$.
In particular, we will discriminate between stars and galaxies on the basis of whether our predicted evaluation of $f$ is positive or negative. 
We will further assume that evaluations of $f$ are corrupted by noise introduced by the observation and preprocessing of data.

The Bayesian model for the GP prior on $f$ for any finite  $X = \{\mathbf{x}_1, \dots, \mathbf{x}_n \} \subset \mathcal{X}$ is given by
\begin{align}
\mathbf{y} &= \mathbf{f} + \boldsymbol{\epsilon} \label{eq:bayesian_model} \\
\mathbf{f} &= [f(\mathbf{x}_1), \dots, f(\mathbf{x}_n)]^\top \sim \mathcal{N}(\mathbf{0}, K_\mathbf{ff}) \label{eq:prior} \\
\boldsymbol{\epsilon} &\sim \mathcal{N}(0, \sigma^2 I).
\end{align}
Here $\mathbf{y}$ is an $n$-vector of the evaluations of $f$ on $X$.
$\mathcal{N}(\boldsymbol{\mu}, \Sigma)$ denotes the multivariate Gaussian distribution with mean vector $\boldsymbol{\mu}$ and covariance $\Sigma$.
$K_\mathbf{ff}$ is an $n \times n$ matrix whose $(i,j)$th element is $k(\mathbf{x}_i, \mathbf{x}_j) = \text{cov}(f(\mathbf{x}_i), f(\mathbf{x}_j))$.
$\boldsymbol{\varepsilon}$ is homoscedastic Gaussian noise affecting the measurement of $f$.
Such covariance matrices implicitly depend on $\theta$.

Say that we are given \emph{labelled} data $X, \mathbf{y} \subset \mathcal{D}$.
Here $X$ is the set of $n$ PCA-embedded images as described in \autoref{sec:reduction}, and $\mathbf{y}$ is a conforming vector whose entries are $+1$ if the image has been deemed a star, and $-1$ otherwise.  
Now consider the $m$ unlabelled embedded images $X_* = \{\mathbf{x}_1^*, \dots, \mathbf{x}_{m}^* \} \subset \mathcal{X}$.
The joint distribution of $\mathbf{y}$ and the response $\mathbf{f}_*$ of $f$ applied to $X_*$ is
\begin{equation} \label{eq:joint_distribution}
  \begin{bmatrix} \mathbf{y} \\ \mathbf{f}_*
  \end{bmatrix}
  = \mathcal{N} \left ( 0,
  \begin{bmatrix}
    K_\mathbf{ff} + \sigma^2 I & K_\mathbf{f*} \\
    K_\mathbf{*f} & K_{**}
  \end{bmatrix} 
      \right ).
\end{equation}
Here $K_\mathbf{*f} = K_\mathbf{f*}^\top$ is the cross-covariance matrix between $X_*$ and $X$, i.e. the $(i, j)$th element of $K_\mathbf{*f}$ is $k(\mathbf{x}_i^*, \mathbf{x}_j)$.
Meanwhile, $K_\mathbf{**}$ has $(i, j)$th element $k(\mathbf{x}_i^*, \mathbf{x}_j^*)$.
Applying Bayes rule to \eqref{eq:prior} and \eqref{eq:joint_distribution} allows us to analytically compute the posterior distribution of the response $\mathbf{f}_*$ on $X_*$ as
\begin{align}
  \mathbf{f}_* \mid X, X_*, \mathbf{y} &\sim \mathcal{N}(\bar{\mathbf{f}}_*, C) \\
  \bar{\mathbf{f}}_* &\equiv K_\mathbf{*f} Q_\mathbf{ff}^{-1} \mathbf{y} \\ 
  C &\equiv K_{**} - K_\mathbf{*f} Q_\mathbf{ff}^{-1} K_\mathbf{f*} \\
  Q_\mathbf{ff} &\equiv K_\mathbf{ff} + \sigma^2 I. 
\end{align}

We use a simple discriminator on the posterior of the response on $\mathbf{X_*}$: if $\bar{\mathbf{f}}^*_i \geq 0$, $x^*_i$ is declared a star, and a galaxy otherwise.
Conveniently, we can examine the $i$th diagonal entry of $C$ to quantify the variance of this prediction.
This allows us to automatically detect when the posterior variance is large, and the model is not confident in its prediction. 

In this paper, we use the Radial Basis Function (RBF) kernel which is a very common choice for a wide range of applications. It is expressed as:
\begin{equation} \label{eq:rbf}
  k_{\textrm{RBF}}(\mathbf{x}, \mathbf{x}^\prime) = \text{exp}\left ( -\frac{\|\mathbf{x} - \mathbf{x}^\prime\|_2^2}{\ell^2} \right ),
\end{equation}
where the only hyperparameter $\ell$ is the length scale of the kernel. Hyperparameter estimation of $\ell$ is achieved by maximizing prediction accuracy through local leave-one-out cross validation. $\sigma^2$ is fixed at the nominal value of 0.015. 

\section{Results}
\label{sec:results}

\begin{figure}[!h]
  \includegraphics[width=\linewidth]{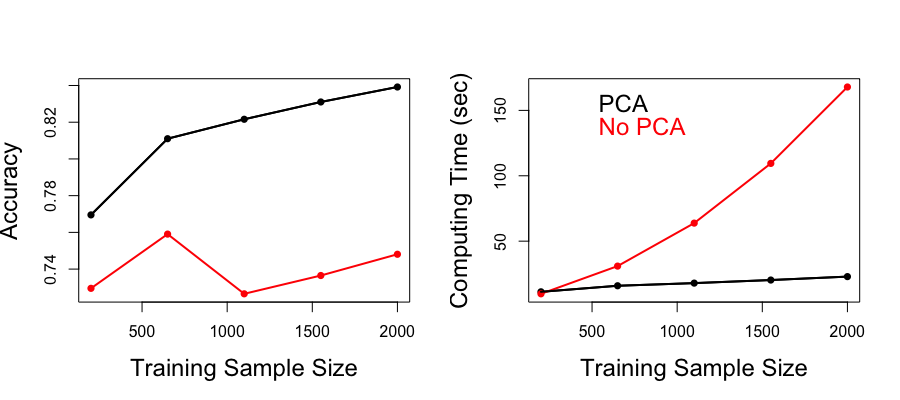}
  \caption{Comparison of prediction accuracy and computing time as a function of training sample size with (black) and without (red). Reducing the dimensionality of the model before applying Gaussian Processes yields a better prediction accuracy and scales much better as the training sample size increases.}
  \label{fig:pca}
\end{figure}

Our main result is the vast performance increase, both in terms of accuracy and computing time, when working with a reduced-order training data set vs the full-dimensional one. 
\autoref{fig:pca} shows that, unsurprisingly given the almost 800-fold reduction in dimensionality, processing time is much lower and scales much better after a PCA reduction, and perhaps more surprisingly, that the prediction accuracy also increases substantially. In particular, over the training sample sizes tested, there is not a significant pattern of increasing mean accuracy with increased training size in the naive method. In contrast, the mean accuracy of our method with data reduction performs better with 200 observations than the naive method with 2000 observations, and further improves as the training size increases. In terms of computing time, including an approximate PCA computation, with a training size of 2000, we reduce the computing time by approximately 86.4\%. These results can be seen in \autoref{fig:pca}.

\begin{figure}[!h]
  \includegraphics[width=\linewidth]{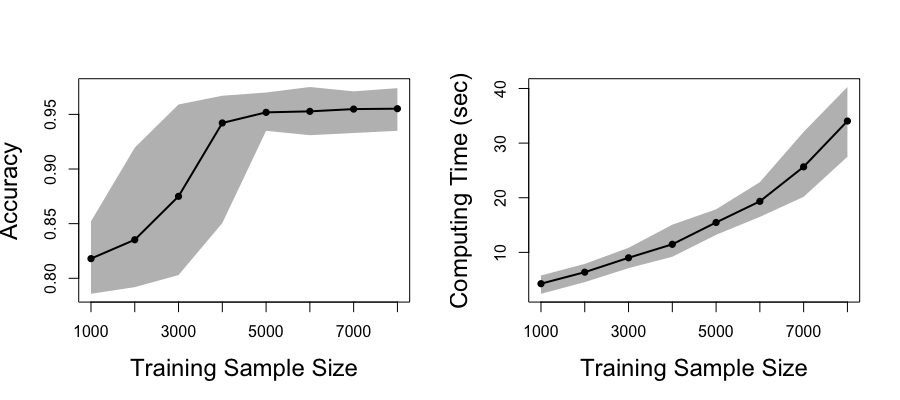}
  \caption{Prediction accuracy and computing time as a function of training sample size for the RBF kernel for large training sample sizes. The shaded region is the empirical simulation 94\% prediction interval of performance. The prediction accuracy starts plateauing after 5000 samples and reaches a maximum accuracy of 0.96.}
  \label{fig:rbf}
\end{figure}

We also explored the accuracy improvement of our methods for larger data sizes in \autoref{fig:rbf}. We see the mean accuracy improves when the number of training samples increases, but this increase is small when once the training sample size reaches 5000 (0.003 change for increase in training size from 5000 to 8000).
At 5000 training samples, the accuracy is 0.96 and the processing time is about 15 seconds.

\section{Conclusion}

We have shown that coupling a simple GP with an RBF kernel with a dimension-reducing embedding using approximate PCA drastically improves the prediction accuracy and computational efficiency of performing star-galaxy separation.
Not only does computational overhead grow more slowly with increasing data scale, but also our prediction accuracy is higher with fewer training examples.
By contrast, DNNs are highly parametric, often utilizing more parameters than observations.
For this reason, DNNs often require large amounts of training data whereas the present method, which also has the advantage of providing full posterior distributions, is effective with only a modest amount of labelled data.

In this paper we chose to use the RBF kernel as it is a common choice in the literature for many applications.
However, the expressiveness of a GP is heavily dependent upon the choice of kernel function.
We suspect that significantly improved accuracy is possible by utilizing more expressive kernel functions.
Furthermore, we have yet to exploit the GP posterior distributions to perform uncertainty quantification in this application.
These investigations are to be the subjects of future work.

\section*{Acknowledgments}
This work was performed under the auspices of the U.S. Department of Energy by Lawrence Livermore National Laboratory under Contract DE-AC52-07NA27344.
Funding for this work was provided by LLNL Laboratory Directed Research and Development grant 19-SI-004.

\bibliographystyle{plainnat}
\bibliography{Star-Galaxy}

\end{document}